# The Role of Metadata in Non-Fungible Tokens: Marketplace Analysis and Collection Organization

A pre-print
Sarah Barrington
U.C. Berkeley, School of Information
December 2021
sbarrington@berkeley.edu

## 1. ABSTRACT

An explosion of interest in Non-Fungible Tokens (NFTs) has led to the emergence of vibrant online marketplaces that enable users to buy, sell and create digital assets. Largely considered contractual representations of digital artworks, NFTs allow ownership and authenticity to be proven through storing an asset and its associated metadata on a Blockchain. Yet, variation exists between chains, token protocols (such as the ERC-721 NFT standard) and marketplaces, leading to inconsistencies in the definitions and roles of token metadata. This research thus aims to define metadata in the context of NFTs, explore the boundary of metadata and asset data within tokens, and understand the variances and impacts these structures have on the curation of NFTs within online marketplaces and collections.

## I. INTRODUCTION

Cryptocurrency and Blockchain technologies are fast becoming areas of public interest across a breadth of diverse domains, with novel applications ranging from financial services[1] to state politics[2]. In particular, promising future use cases of these decentralized technologies aim to re-define the concept of value, and potentially enable a new wave of accessible investment opportunities for the general public.

One such example of this is through Non-Fungible Tokens (NFTs), which act as digital certificates of ownership for a given piece of digital content[3]. These stamps are stored as tokens on a public Blockchain (such as Ethereum), which uses complex hashing algorithms to ensure that each token is unique, tamper-proof and publicly-visible. This adds value to a range of digital content & artworks through enabling a single source of truth relating to its ownership and historic activity.

In particular, a vibrant online market is emerging in response to NFT technologies, with multiple web-based marketplaces being created to meet the demand of a growing ecosystem of creators, buyers and sellers. These websites focus on speed and usability; enabling users to create, buy and sell NFTs within minutes. However, the nascence of this technology means that associated language and conceptual understanding of the process remains somewhat esoteric; and the data that is both required and created through this process can still act as a barrier to entry for new users who are not familiar with the fundamental concepts of decentralized technologies.

The aim of this research paper is thus to answer the following research questions:

- What metadata is required, or created, in the process of creating an NFT?
- What is the boundary between the 'metadata' and 'data' of an NFT?
- How do popular online marketplaces store and communicate this metadata?
- How does this impact the curation of online NFT collections?

This article will address these research questions through a brief review summarizing the technical architecture of data and metadata in NFTs, examining NFT listings on 2-3 leading marketplaces and performing a general collection analysis of the

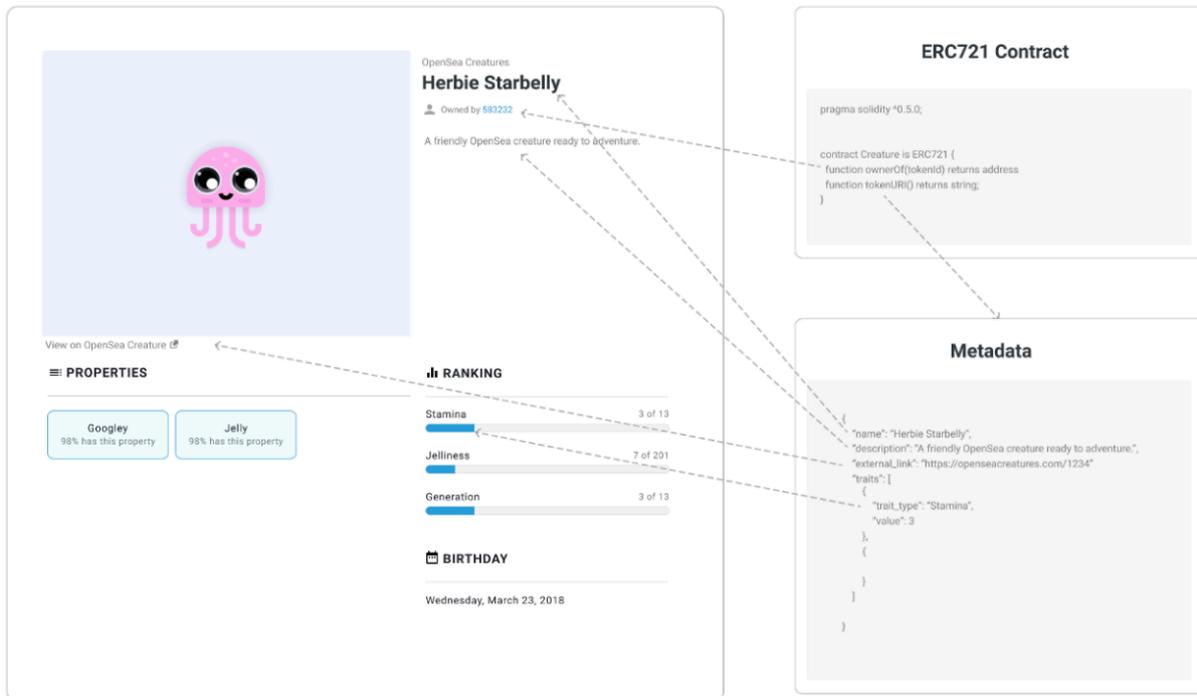

Fig 1. Example of NFT containing both ERC721 Contract and associated Metadata. *Devin Finzer, OpenSea Blog post, Section 3 Non-Fungible Token Metadata, January 10 2020*

observed approaches by which online NFT collections are curated.

## II. ANALYSIS OF METADATA

Metadata is largely considered to be data that provides information about other data[4]. In practice, the term is most frequently used to distinguish the difference between an object, dataset or piece of information (the 'data'), and it's associated descriptors and content attributes (the 'metadata'). Metadata can take many forms, including descriptive, structural[5] and administrative[6], all of which can be used to support a variety of information activities ranging from resource discovery and identification to collection organization[7].

Despite the breadth of formats metadata can assume, several standards have emerged that formalize the set of collected attributes for specific types of information, including the Dublin Core Metadata Initiative[8] for internet resources and The Text Encoding Initiative (TEI)[9] for electronic texts. In rapidly-evolving modern day computing architecture, new and more complex forms of information are emerging that expand beyond existing standards, with NFTs in particular presenting a novel context in which metadata plays an important role.

### Application of Metadata in ERC721 Standard Tokens

Blockchain ledgers are used widely for the transfer of fungible (i.e. interchangeable) tokens, including cryptocurrencies such as Bitcoin and Ethereum. The introduction of Non-Fungible Tokens, and related concepts of digital scarcity and authenticated ownership, was pioneered through the ERC721 Ethereum standard lead by digital collectibles company CryptoKitties. This standard is a systematized interface for defining and interacting with a digital asset comprising two methods stored in an ethereum-based Solidity smart-contract: 1) **ownerOf** and 2) **transferFrom**[10]. Solidity is an object-oriented language for implementing smart contracts[11].

It is the combination of these two methods that comprise the representation of an NFT; defining a transaction mechanism alongside a current owner. Public users can write Ethereum queries using each method to deduct information about token provenance and ownership. However, these two standard methods do not return the digital asset itself, nor its associated attributes. The introduction of **NFT metadata** to the architecture is thus required to couple the smart contract proof-of-

```json
{
  "name": "Duke Khanplum",
  "image":
"https://storage.googleapis.com/ck-kitty-
image/0x06012c8cf97bead5deae237070f9587f8e7a2
66d/1500718.png",
  "description": "Heya. My name is Duke
Khanplum, but I've always believed I'm King
Henry VIII reincarnated."
}
```

Fig 2. Example of Off-Chain NFT JSON dictionary containing asset metadata - both attributes, and link to the asset itself [as footnote 13]

ownership with the information it pertains to. The ERC721 standard thus ensures every NFT is globally and uniquely identifiable, transferrable and can optionally include metadata[12].

### Conceptual Definition of NFT Metadata

Thus, the traditional definition of metadata is somewhat specialized in the case of NFTs, as the term includes **both** the **attributes** and **the original information** itself. This could be argued to be reflective of how decentralized web architects consider the NFT infrastructure; in which the core data is the smart-contract itself that exists on the Ethereum blockchain, and the underlying digital asset is viewed as auxiliary to this. This is true of both possible scenarios in which this metadata is stored directly on the blockchain ledger, or elsewhere. This difference is illustrated in Fig 1., in which the ERC721 smart-contract indicating ownership is displayed separately from the information regarding the asset itself.

### On-Chain Metadata

When NFT metadata exists on-chain, the metadata (including both the underlying digital asset and it's relevant descriptor attributes) is built directly into the smart-contract[13]. This enables the core benefits of true asset permanence, as Blockchains are immutable and thus the inclusion of an asset within the original smart-contract guarantees scarcity and long-lasting existence of the asset. In particular, this ensures the digital asset exists if, for example, the original website host or creator server is taken down[14].

### Off-Chain Metadata

However, on-chain metadata presents storage issues for the current Ethereum infrastructure. To account for this, the ERC721 standard was adapted to include a link function (tokenURI) that points the smart-contract towards a public URL that returns a JSON dictionary of data containing both the Digital asset link and it's assocaited metadata[15], as can be seen in Fig 2.

In this example, the URL provided in the "image" metadata attribute pertains to the digital location of the underlying asset, which can be stored in a variety of ways ranging from centralized cloud servers to the InterPlanetary File System (IPFS), an emerging new peer-to-peer file storage technology.

### III. MARKETPLACE ANALYSIS

In order to facilitate the creation, marketing and transferral of ERC721 NFTs, platforms offer a range of models by which to tailor smart-contracts and their associated metadata for a specific use case or user group. In particular, these differing areas of focus can be observed across three of the current market-leading NFT marketplaces: OpenSea, Rarible and SuperRare.

### OpenSea

OpenSea is presently the world's largest NFT marketplace, surpassing $10.35bn (USD) in sales since its inception in 2017. The site operates as both a website and an app, and enables users to create, share and trade Ethereum and other blockchain-supported NFTs through a simplified user interface. NFT creators can build and monetize both on-chain (e.g. OnChainMonkey[16]) and off-chain NFTs, and each creator can organize these into **collections** according to their metadata.

### Metadata Creation Process

In particular, OpenSea facilitates a wide range of customizable metadata options for the creation of each NFT that can be included into the smart contract - removing the onus from users to manually program this directly into a solidity interface.

| Contract Address | 0xab0b...0879 |
|---|---|
| Token ID | 2396 |
| Token Standard | ERC-721 |
| Blockchain | Ethereum |

Fig 3. Example unique contract address and token ID for an NFT on OpenSea [link]

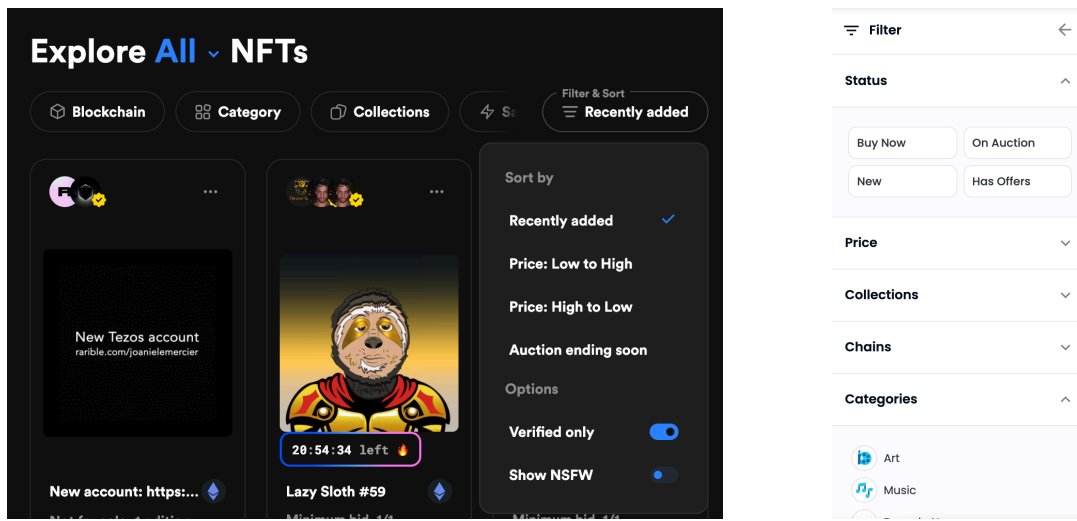

Fig 4. Search interfaces with faceted categories for Rarible (left) and OpenSea (right)

Through recognizing the need for flexibility and richness in defining metadata, OpenSea has created a vibrant and diverse digital asset marketplace that is accessible to a wide range of skillsets.

## Rarible

Rarible is a smaller-scale marketplace reporting a sales total of $274m (USD) since it was founded in 2020. As with the majority of platforms, users are required to link an existing Ethereum wallet (a way of representing an account to interact with the blockchain), and then Rarible acts as the interface between the user and the blockchain architecture.

### Metadata Creation for Supporting Collections

In particular, Rarible describes its user collections as analogous to 'a folder on [a] desktop'[17]. When artwork in the form of an NFT is created, or 'minted', it is placed into the appropriate public collection 'folder' according to it's relevant standard (either ERC721, or ERC1155, am adaptation of the ERC721 standard that enable the creation of various token types). When users create their own collection, this can be considered a 'sub-folder' within the broader token folder, indexed according to the creator or owner.

The creation of a sub-collection requires the deployment of an additional smart-contract that associates a given user with each NFT they create or own, in addition to the default association with the overarching Rarible collection[18]. This further augments the metadata associated with a given NFT, in which Rarible implements the additional 'membership' smart-contract on behalf of users wishing to create their own sub-collections.

## SuperRare

SuperRare is a hybrid social-marketplace that focusses on creating, curating and transacting high value digital collectibles on behalf of users. Similarly to OpenSea and Variable, SuperRare provides a simplified interface between the user and the back-end ERC721 smart-contract & metadata.

One key difference to OpenSea and Rarible is that creators must apply and subsequently be accepted before being on-boarded to the platform. SuperRare thus places focus on exclusivity and scarcity in both the technical and artistic senses, and enables *secondary market revenue;* a unique property of NFTs that allows the original artist to automatically retain profits each time their artwork is re-sold. Only ERC721 tokens are currently supported.

### Metadata for Social Networking

A further core value proposition of SuperRare is the focus on creating a social network around the base NFT functionality of buying, selling and transacting. Although Rarible and OpenSea facilitate browsing and the creation of collections; SuperRare aims to connect users with each other through leveraging the public nature of each NFTs metadata. One feature supporting this is through the metadata displayed in the market area of the site, in which users can browse pieces according to both the owner and the original artist. Not only does this display of prior ownership metadata allow clear traceability of primary and secondary royalties, but it ensures the creator gets credit and ultimately equal exposure to traffic in aid of building a social presence on the site.

## Collection Analysis

In all three cases, the NFTs that users have created on the platform are arranged into collections that resemble those of the physical art world. Each marketplace places focus on differing aspects of NFT metadata depending upon user group and value proposition of the platform. However, commonalities can be observed in the way in which metadata is used to support users and creators in navigating, browsing and curating collections.

### Organizing Strategy

In all cases, interaction with NFTs in any collection can only be **accessed** through creating a user account and associating a pre-existing blockchain *wallet*; a form of digital identity that establishes users as formal agents in the blockchain ledger and enables them to hold and transact tokens.

Each site has an explore or discover feed, that acts as a starting point for non-targeted browsing, alongside a specific search bar functionality. NFTs are arranged into sub-collections that typically centre upon a given creator, and new NFTs can typically be sold for a fixed price or opened as an auction.

Within any given collection, the NFTs are arranged as single instances of digital art, each with a single smart-contract and associated dictionary of metadata. As the tokens are inherently non-fungible, resources are **not interchangeable**. The **resource granularity** is thus defined at the singular token level- with the exception being for ERC1155 tokens, in which a single resource pertains to a set of digital assets (although bound by a singular token). Each are **identified** by their unique smart contract address and token ID (as shown in Fig. 3), all of which are permanent and unique on the blockchain.

The **organizing strategy** in most cases is limited by the meta data associated with the NFT, and so centers around the creator or owner. Creating more complex collections involves either manual tagging of the asset for website-specific use only, or deploying further smart contract wrappers to augment the initial limited functionality (as was the case with deploying a sub-collection on Rarible).

The **selection criteria** for each NFT is that the token standard is compatible with that of the platform (all ERC721 in the above examples), alongside an association with both a creator/owner and the platform itself (for example, ensuring NFTs minted on Rarible are not cross-listed on OpenSea because the Ethereum blockchain is public and technically both are visible).

**Interactions** with collections occur through users searching for a specific artist or collection name, examining the listing with it's associated metadata and website-based text descriptions, and browsing for art to purchase or transact with the aid of **faceted metadata-based categories.**

### Faceted Metadata

Faceted categories allow users to navigate explicitly along conceptual dimensions that describe the resources within a collection[19]. In particular, faceted categories are widely used in online browsing, enabling users to 'filter' on various attributes of interest or exclusion, in order to (often iteratively) narrow the search space.

For both OpenSea and Rarible, the faceted categories by which users could browse for collections and NFTs share commonalities (Fig 4.). These include applying filters according to price, collection (by owner) and the hosting blockchain (primarily Ethereum, but with some optionality for Polygon and Klatyn on OpenSea, and Flow and Tezos for Rarible). It was observed that every faceted category that could be used to filter was flat- I.e. without a hierarchical structure of other categories associated to it.

### Usability and the Role of Metadata in Search

These categories, or 'facets', pertain to a mixture of formal NFT metadata (including collection by owner and chain) and marketplace-assigned metadata (such as price and auction vs. Open sale). Without background in understanding the boundaries of NFT metadata, and how it can be tailored by each platform in different ways, less experienced users may initially find the filter options somewhat constrained and inflexible due to the limited number of facets available to select from and their technical specificity.

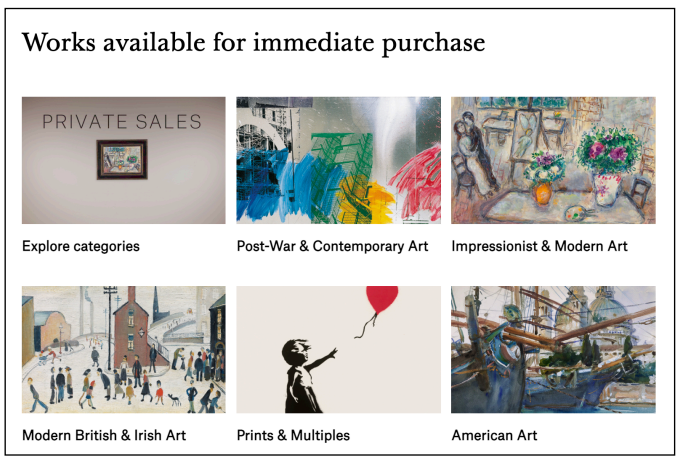

Fig 5. The private sale artworks collection from Christies.com, with metadata arranged so as to enable navigation by hierarchy as opposed to faceted category (https://www.christies.com/)

For example, introducing a further level of hierarchy into the 'categories' facet on OpenSea might enable more richness in the browsing process, allowing users to drill down into more specific areas of interest within broader categories such as 'collectibles' and 'utility'. Although these these terms are likely well understood by experienced users and early blockchain technology adopters, newer users may not immediately appreciate the relevance of them to a piece of digital artwork.

This extends to the inclusion of the 'chain' facet; by which users can select which Blockchain they wish the NFTs in their search to be developed upon. These observations reflect both the technical constraints of metadata, but also allow for deductions about likely user interest and behaviors in searching for NFTs- that the focus is largely on the technology and infrastructure, and far less on the content of the digital artwork itself.

SuperRare self-brands as 'Instagram meets Christies'[20], referring to the social media network alongside the traditional fine art & antiquities auction house. Drawing a comparison between the search experience of SuperRare and Christies online private sales collection[21], it can be observed that Christies makes use of a hierarchical navigation structure centering upon the artwork itself, as opposed to the method of sale, current or previous ownership or mechanism of creation. It is expected that there will be differences in browsing for real-world and digital artworks, however, the browsing experience of Christies makes for a far more immersive experience and allow even inexperienced users to navigate by more nuanced and visual means (Fig 5.).

## IV. CONCLUSIONS

This research aimed to review the current technical architecture of NFT metadata, alongside outlining how it is distinct from the digital asset itself. An overview of three leading online marketplaces revealed three nuanced models for tailoring NFT metadata in order to serve different use cases and groups of customers. Performing a collection analysis for the commonalities exhibited between the platforms demonstrated the value of using metadata for organization strategy; and that the non-fungibility of these tokens allows streamlined curation at the resource level.

Finally, the above observations set the foundation for the suggestion that metadata in NFTs provides great technical and authentication value; yet may be insufficient alone for supporting user-led search and navigation queries. Future work could focus upon two areas of interest:

- What further manual (non-metadata) labels could marketplaces assign to NFTs to support a richer browsing experience?
- How could these inform better faceted categories, or a less-esoteric user experience for new users seeking to gain exposure to the industry?

Although NFT marketplaces are becoming increasingly commonplace, there is work to be done in order to bridge the technical appeal with the artistic value of each piece.